\documentclass[twocolumn]{jpsj2} 

\usepackage{amsmath,psfrag,graphicx,times}

\def\dd{{\mathrm d}}
\def\ii{{\mathrm i}}
\def\ee{{\mathrm e}}

\def\ket#1{|#1\rangle}
\def\bra#1{\langle #1|}

\def\bracketii#1#2#3{\langle #1|#2|#3 \rangle}

\def\ha{\hat{a}}
\def\p{^\prime}
\def\E{\mathcal{E}}
\def\P{\mathcal{P}}
\def\H{\hat{H}}
\def\infint{\int^{\infty}_{-\infty}}
\def\infiint{\int^{\infty}_{-\infty}\int^{\infty}_{-\infty}}

\begin{document}

\title{Full Quantum Analysis of Two-Photon Absorption
Using Two-Photon Wavefunction: Comparison with One-Photon Absorption}

\author{Toshihiro Nakanishi$^{1,2}\thanks{E-mail: t-naka@kuee.kyoto-u.ac.jp}$,
Hirokazu Kobayashi$^{1}$,\\
Kazuhiko Sugiyama$^{1,2}$,
and Masao Kitano$^{1,2}$}

\inst{Department of Electronic Science and Engineering,
Kyoto University, Kyoto 606-8501, Japan$^1$\\
CREST, Japan Science and Technology Agency, Saitama 332-0012, Japan$^2$}

\abst{
For dissipation-free photon-photon interaction at the single photon level,
we analyze one-photon transition
and two-photon transition induced by photon pairs
in  three-level atoms
using  two-photon wavefunctions.
We show that the two-photon absorption can be substantially enhanced 
by adjusting the time correlation of photon pairs.
We study two typical cases: Gaussian wavefunction and
rectangular wavefunction.
In the latter,
we find that under special conditions  one-photon transition
is completely suppressed while the high probability of 
two-photon transition is maintained.
}

\kword{
two-photon absorption, photon pair, quantum optics,
two-photon wavefunction, time correlation, entanglement-induced
two-photon transparency
}

\maketitle

\section{Introduction}

Two-photon absorption is one of the most fundamental 
nonlinear processes,
and it sometimes reveals the quantum nature of light \cite{Loudon:OC84}.
If a two-photon transition  is
achieved at a single-photon level,
it can be possible to
control the fate of a single photon by controlling the
presence or absence of the other single photon.
Such photon-photon interaction may be applicable
to quantum information technologies,
such as the development of  optical switches for two photons
and Bell-state analyzers \cite{Tomita:01} among others.
\cite{Franson2004, PhysRevLett.83.3558, PhysRevLett.85.2733,
Nakanishi:PRA03,Franson2006,Jacobs2006}.
In these applications,
the medium must be such that it absorbs the 
 two photons and not one of them.
In three-level systems having a ground state, intermediate state,
and an excited state, two-photon absorption and 
one-photon absorption  to the intermediate level can occur 
simultaneously.
The two-photon absorption can be enhanced by
decreasing the detuning to the intermediate state;
however, this may also result in an increase in the
 one-photon transitions to the intermediate level.
Several methods for enhancing the two-photon absorption
while suppressing the one-photon absorption
have been proposed.
The use of 
electromagnetically induced transparency (EIT) is one approach
for realizing  the suppression of  one-photon transition
while maintaining the strong nonlinearity
in four-level atoms using auxiliary light called the coupling light
\cite{Schmidt:ol96,PhysRevLett.77.1039,kang:093601}.
The cavity QED is another promising method for obtaining effective
two-photon absorption \cite{turchette:4710}.

It is also possible to enhance two-photon absorption by
tailoring the quantum state of photons \cite{Javanaien:PRA90,fei:1679};
this does not involve the use of
any external apparatus such as a cavity or a control light,
as in the case of the EIT system.
To obtain efficient two-photon excitations,
the two photons must satisfy the following two conditions:
they should be close to each other, and
the total linewidth of the two photons (two-photon linewidth)
must be less than the linewidth of the excited state.
Satisfying the first condition enables the occurrence of
instantaneous transition via the virtual state, 
and the second one ensures that two-photon resonance occurs for a long time.
An ultrashort light pulse does not satisfy the second condition
because of its broad spectrum.
In contrast, twin photons produced 
by spontaneous parametric down-conversion (SPDC) processes
using a continuous pump laser
satisfy both the conditions
and can effectively undergo two-photon transitions.
However, the suppression of the 
one-photon transitions to the intermediate state
and  enhancement of the two-photon transitions
have not been investigated in detail.

In order to evaluate the transition probabilities
of photon pairs,
full quantum treatment of the system is required.
We consider a propagating light beam consisting of continuous modes,
which requires a multimode description.
To carry out the full quantum multimode analysis of this light beam,
we introduce two-photon wavefunctions that can
represent arbitrary two-photon states 
 \cite{PhysRevA.42.4102,Scully,Shih:IEEE03,
PhysRevLett.93.173601,koshino:013804}. 
Our formulation allows us to analytically estimate
not only two-photon absorption
but also the one-photon absorption.

In Sec.~\ref{Sec:Theory},
we describe the formulation  both  the two-photon  and one-photon absorption
for arbitrary two-photon states using two-photon wavefunction.
In Sec.~\ref{Sec:pairs}, we compare the 
two-photon  and  one-photon absorptions
for two types of time-correlated photon pairs.
Finally, we show that by adjusting the correlation time,
the single-photon loss can be suppressed
without decreasing the high probability of two-photon absorption.

\section{Theory\label{Sec:Theory}}

\subsection{Field operators in free space}
\begin{figure}[]
\begin{center}
 \psfrag{wi}[c][c]{$\omega_i$}
 \psfrag{w}[c][c]{$\omega$}
 \psfrag{W}[c][c]{$\Omega$}
 \psfrag{Wi}[c][c]{$\Omega_i$}
 \psfrag{W1}[c][c]{$\Omega_a$}
 \psfrag{W2}[c][c]{$\Omega_b$}
 \psfrag{w1}[c][c]{$\omega_1$}
 \psfrag{w2}[c][c]{$\omega_2$}
 \psfrag{g}[c][c]{$\ket{g}$}
 \psfrag{a}[c][c]{$\ket{a}$}
 \psfrag{b}[c][c]{$\ket{b}$}
 \psfrag{e}[c][c]{$\ket{e}$}
 \includegraphics[width=7cm]{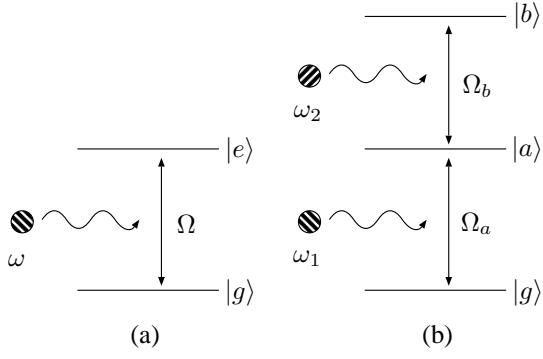}
 \caption{Level diagrams for (a) one-photon transition and (b) 
two-photon transition.}
 \label{Fig:level}
\end{center}
\end{figure}

In order to investigate the interaction of atoms with an
electromagnetic field propagating in one dimension,
we define the electric field in continuous modes as
\begin{align}
 \hat{E}(z, t) = \infint \dd \omega \sqrt{\frac{\hbar \omega}{4 \pi \epsilon_0
 c A}} \ha(\omega) \, \ee^{- \ii \omega (t - z/c)} + {\rm H.c.}, 
 \label{Eq:quantization1}
\end{align}
where $A$ denotes the cross section of the beam \cite{PhysRevA.42.4102}.
The annihilation operator $\ha (\omega)$ satisfies the following
commutation relation:
\begin{align}
 [ \ha(\omega), \ha^\dag(\omega\p) ] = \delta (\omega-\omega\p).
\end{align}
If the bandwidth of the field is assumed to be
 considerably less than the carrier frequency
$\bar\omega$, we can simplify eq.~(\ref{Eq:quantization1}) as
\begin{align}
 \hat{E}(z, t) &= \E \, \ha(t-z/c) + {\rm H.c.},
 \label{Eq:quantization2}
\end{align}
where
\begin{align}
 \ha (t) & \equiv
 \frac{1}{\sqrt{2 \pi}} \infint \dd \omega \, \ha(\omega) 
 \, \ee^{-\ii \omega t}, \label{Eq:at}\\
 \E & \equiv \sqrt{\frac{\hbar \bar\omega}{2 \epsilon_0 c A}}.
\end{align}
The Fourier transformed operator $\ha(t)$
has the same commutation relation as $\ha(\omega)$:
\begin{align}
 [ \ha(t), \ha^\dag(t\p) ] = \delta (t-t\p).
\end{align}

\subsection{Single-photon transition}

Before discussing the two-photon transitions, 
we first consider
 the interaction between a wavepacket containing a single photon
and a two-level atom, as shown in Fig.~\ref{Fig:level}(a).
The initial state of the single-photon wavepacket is represented
 as $\ket{\psi}$,
and the atom at $z=0$ is prepared in the ground state $\ket{g}$.
We introduce a simplified notation $\ket{g, \psi}=\ket{g}\ket{\psi}$ 
to represent the composite system.
The Hamiltonian in the interaction picture is
\begin{align}
 \H (t) = \hat{p} (t) \hat{E} (t), \label{Eq:H1}
\end{align}
where 
\begin{align}
 \hat{p} (t) &= \P \ket{e}\bra{g} \ee^{\ii \Omega t} + {\rm H.c.} 
 \label{Eq:pa}
\end{align}
represents an atomic dipole oscillating at the transition frequency $\Omega$
and the electric-dipole transition matrix element  given as
$\P=-e\bracketii{e}{r}{g}$
\cite{Scully}.

By applying the first-order perturbation theory,
the probability amplitude for the atom to exist in the excited state $\ket{e}$
after the passage of the wavepacket is given as
\begin{align}
 \alpha_1 = -  \frac{\ii}{\hbar} \infint \dd t \bra{e, 0} \H(t)
 \ket{g, \psi},
 \label{Eq:alpha_1}
\end{align}
where $\ket{e, 0}=\ket{e}\ket{0}$ with the vacuum state $\ket{0}$.
We assume that the transit time of the wavepacket is shorter than
the relaxation time of the excited state $\ket{e}$.
On substituting eqs.~(\ref{Eq:quantization2}) and 
(\ref{Eq:H1}) in eq.~(\ref{Eq:alpha_1}),
we obtain
\begin{align}
 \alpha_1 = -  \frac{\ii \E \P}{\hbar} \infint \dd t \,
 \ee^{\ii \Omega t} \bracketii{0}{\ha(t)}{\psi}.
\end{align}
In this derivation, we use the rotating wave approximation.
We introduce the one-photon amplitude 
\begin{align}
 \psi (t) \equiv \bracketii{0}{\ha(t)}{\psi}, \label{Eq:one-wf}
\end{align}
whose square is proportional to the probability
of photon detection at time $t$ \cite{Scully}.
Its Fourier transform is expressed as
\begin{align}
 \Psi(\omega) = \frac{1}{\sqrt{2 \pi}} \infint \dd t \, \psi(t) \,
 \ee^{\ii \omega t},
\end{align}
the square of which represents
the spectral intensity of the wavepacket;
then,
the probability of  excitation, $P_1\equiv|\alpha_1|^2$, 
can be expressed as
\begin{align}
 P_1 &= 2 \pi r |\Psi(\Omega)|^2, \label{Eq:one}
\end{align}
where $r \equiv \E^2 \P^2 / \hbar^2$.
Thus, the probability of
the one-photon absorption is determined by
the spectral component $|\Psi(\Omega)|^2$
at the atomic transition frequency of $\Omega$.

\subsection{Two-photon transition}

In this section, we discuss the two-photon transitions induced by
a pair of photons, as shown in Fig.~\ref{Fig:level} (b).
Let us consider that photon 1 in a mode induces the lower transition,
 $\ket{g} \rightarrow \ket{a}$ while
photon 2 in another mode is responsible for
the upper transition, $\ket{a} \rightarrow \ket{b}$.

The interaction Hamiltonian for this scenario 
is represented as
\begin{align}
 \H(t) &= \hat{p}_1(t) \hat{E}_1(t) + \hat{p}_2(t) \hat{E}_2(t),
\end{align}
where
\begin{align}
 \hat{p}_1(t) &= \P_a \ket{a}\bra{g} \ee^{\ii \Omega_a t} + {\rm H.c.},\\
 \hat{p}_2(t) &= \P_b \ket{b}\bra{a} \ee^{\ii \Omega_b t} + {\rm H.c.}
\end{align}
To estimate the two-photon excitation probability,
we use the second-order perturbation theory.
The second-order component of the time evolution operator
is expressed as
\begin{align}
 \hat{U}_2 &= - \frac{1}{\hbar^2} \infiint \dd t_2 \dd t_1 
 \H(t_2) \H(t_1) \, \theta(t_2-t_1),
\end{align}
where $\theta(t)$ is the Heaviside step function:
\begin{align}
  \theta(t) = 
\begin{cases}
 1 & (t \geq 0) ,\\
 0  & (t<0).
\end{cases}
\end{align}
When a wavepacket  $\ket{\psi}$, containing a pair of photons,
passes through an atom in the ground state $\ket{g}$,
the probability amplitude of the two-photon excitation can be given as
\begin{align}
 \alpha_2 &= \bracketii{b, 0}{\hat{U}_2}{g, \psi} \nonumber  \\
   &= - \frac{\E_1 \E_2 \P_a \P_b}{\hbar^2} \infiint \dd t_2 \dd t_1
 \ee^{\ii \Omega_b t_2} \ee^{\ii \Omega_a t_1} \nonumber \\
 &\hspace{1.0cm}
 \times \bracketii{0}{\ha_2(t_2) \ha_1(t_1)}{\psi} \, \theta(t_2-t_1).
 \label{Eq:p2}
\end{align}
Similar to the case of eq.~(\ref{Eq:one-wf}),
we can define the two-photon amplitude as
\begin{align}
 \psi (t_1, t_2) = \bracketii{0}{\ha_2(t_2) \ha_1(t_1)}{\psi},
\end{align}
whose square corresponds to the joint probability of finding
 photon 1 at $t=t_1$  and photon 2 at $t=t_2$.
The above two-photon amplitude
 is called {\it an effective two-photon wavefunction}
or {\it a biphoton} \cite{PhysRevA.50.5122, Shih:IEEE03}.
It is clear that the part of $\psi(t_1, t_2)$ 
for $t_2<t_1$
does not contribute to the two-photon excitation
because the absorption of photon 1
is always followed by that of photon 2.
In addition to the Fourier transform of $\psi (t_1, t_2)$, which is
 given as
\begin{align}
 \Psi(\omega_1, \omega_2) &\equiv \frac{1}{2 \pi} \infiint \dd t_2 \, \dd t_1 \,
 \nonumber \\ 
  &  \times \ee^{\ii \omega_2 t_2}\ee^{\ii \omega_1 t_1} \psi (t_1, t_2),
 \label{Eq:Psi}
\end{align}
we also introduce the Fourier transform of $\psi (t_1, t_2) \theta(t_2-t_1)$:
\begin{align}
 \tilde{\Psi}(\omega_1, \omega_2) &\equiv \frac{1}{2 \pi}
 \infiint \dd t_2 \, \dd t_1 \, \nonumber \\
 & \ \times \ee^{\ii \omega_2 t_2}\ee^{\ii \omega_1 t_1} \psi (t_1, t_2)
 \theta(t_2-t_1); \label{Eq:Psi_t}
\end{align}
this Fourier transform represents the spectral intensity of the photon pair
under the time ordering $t_2>t_1$.
From eq.~(\ref{Eq:p2}), we derive the probability of 
two-photon excitation, $P_2=|\alpha_2|^2$, as
\begin{align}
 P_2 = 4 \pi^2 r_1 r_2 |\tilde{\Psi}(\Omega_a, \Omega_b)|^2,
 \label{Eq:P2}
\end{align} 
where
$r_1 \equiv \E_1^2 \P_a^2/\hbar^2$ and $r_2 \equiv \E_2^2 \P_b^2/\hbar^2$.
The probability of  two-photon absorption is given by
the spectral component $|\tilde{\Psi}(\Omega_a, \Omega_b)|^2$
of the time-ordered wavefunction
at the transition frequencies of $\Omega_a$ and $\Omega_b$.

\subsection{One-photon transition induced by one photon of the photon
  pair}

\begin{figure}[]
\begin{center}
\psfrag{omega1}[c][c]{$\omega_1$}
 \psfrag{omega2}[c][c]{$\omega_2$}
 \psfrag{omegaa}[c][c]{$\Omega_a$}
 \psfrag{omegab}[l][l]{$\Omega_b$}
 \psfrag{omegaab}[c][c]{$\bar{\omega}_1$}
 \psfrag{omegabb}[l][l]{$\bar{\omega}_2$}
 \psfrag{delta}[c][l]{$\Delta$}
 \psfrag{tau}[c][l]{$1/\tau$}
 \psfrag{T}[c][l]{$1/T$}
 \psfrag{spectrum1}[l][l]{$|\Psi(\omega_1, \omega_2)|$}
 \psfrag{spectrum2}[l][l]{$|\tilde{\Psi}(\omega_1, \omega_2)|$}
 \includegraphics[width=7cm]{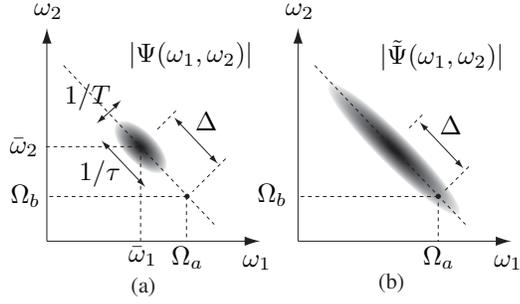}
 \caption{Schematic diagram of (a) $|\Psi(\omega_1, \omega_2)|$
and (b) $|\tilde{\Psi}(\omega_1, \omega_2)|$}
 \label{Fig:spectrum}
\end{center}
\end{figure}

In this section, we consider the probability $P_1$ 
of photon 1 exciting the atom 
to the intermediate level $\ket{a}$.
In eq.~(\ref{Eq:one}), the probability of  one-photon absorption 
is determined by the spectrum of the photon that induces
the transition.
If the two photons are not entangled,
i. e. , $\Psi(\omega_1, \omega_2)=\Psi_1(\omega_1)
\otimes \Psi_2(\omega_2)$,
we can easily derive the excitation probability 
as $P_1 = 2 \pi r_1 |\Psi_1(\Omega_a)|^2$.
However, in the case of an entangled photon,
shown in Fig.~\ref{Fig:spectrum}(a),
the spectrum of photon 1 cannot be represented as the function
of  $\omega_1$ alone,
because the spectrum of photon 1 depends on that of photon 2.
In such cases, the probability of 
one-photon absorption is obtained by tracing over the state of  photon 2 as
\begin{align}
 P_1 = 2 \pi r_1 \infint \dd \omega_2^\prime |\Psi(\Omega_a, \omega_2^\prime)|^2.
 \label{Eq:P1}
\end{align}

\section{Two-photon absorption induced by time-correlated photon pairs
\label{Sec:pairs}}

Before introducing the two-photon wavefuction specific to this study,
we will discuss the general properties of 
two-photon absorption induced by a time-correlated photon pair.
The two-photon wavefunction is characterized by
two parameters:
coherent time $T$ and correlation time $\tau$.
The coherence time $T$ is equal to the coherence time of the pump
field.
The correlation time $\tau \, (\ll T)$ corresponds to the time period
required for both the photons to be detected by the two detectors
and is determined by the spectral width of the down-converted light.
In the frequency domain,
the two-photon wavefunction $\Psi(\omega_1, \omega_2)$ can be represented
as the product of two functions: $\Psi_+(\omega_+)$ with a width 
of $1/T$ and $\Psi_-(\omega_-)$ with a width of $1/\tau$
, where $\omega_+ \equiv \omega_1 + \omega_2$ and
$\omega_- \equiv (\omega_1 - \omega_2)/2$,
as shown in Fig.~\ref{Fig:spectrum}(a).
From eq.~(\ref{Eq:Psi_t}),
$\tilde{\Psi}(\omega_1, \omega_2)$ is represented as
the convolution integral: 
\begin{align}
 \tilde{\Psi}(\omega_1, \omega_2) = 
 \Psi_+(\omega_+) \int \dd \omega_-^\prime \Psi_-(\omega_- -
 \omega_-^\prime) \Theta(\omega_-^\prime),
\end{align}
where $\Theta(\omega)$ is the Fourier integral of 
the Heaviside step function $\theta(t) $ and is given by
\begin{align}
 \Theta(\omega) = 
 \sqrt{\frac{\pi}{2}} \delta(\omega)
 - \frac{\ii }{\sqrt{2 \pi}}\frac{\rm P}{\omega}. \label{Eq:step}
\end{align}
The symbol ${\rm P}$ denotes the Cauchy principal value \cite{Arfken}.
The second term in the above equation causes
the spectrum of  ${\Psi}(\omega_1, \omega_2)$  to broaden
in the direction of the frequency difference $\omega_-$,
as shown in Fig.~\ref{Fig:spectrum}(b),
because of its convolution integral with $1/\omega$.

For simplicity, it is assumed that
\begin{align}
 \Psi_+(\omega_+) &= \frac{\beta}{\sqrt{T}} \phi((\omega_+ 
- \bar{\omega}_+) T) \nonumber \\
 \Psi_-(\omega_-) &= \frac{\beta}{\sqrt{\tau}} \phi((\omega_- - 
\bar{\omega}_-) \tau), \quad
\end{align}
where the common function $\phi(\alpha \omega)$,
which satisfies $\int \phi(\alpha \omega) \dd \omega=1$,
has a width of $1/\alpha$,  height of $\alpha$, and 
shows a peak at $\bar\omega_\pm$.
(The coefficient $\beta$ must be determined to satisfy the
 normalization condition
$\int \dd \omega_\pm |\Psi_\pm (\omega_\pm)|^2=1$.)
When the two-photon-resonance condition is satisfied as
$\bar\omega_+=\Omega_+ (\equiv \Omega_a+\Omega_b)$, 
we get ${\Psi}_+(\Omega_+)=\beta \phi(0)/\sqrt{T}=\beta \sqrt{T}$.
If the detuning $\Delta$ is sufficiently large for
$\phi((\omega_- - \bar\omega_-)\tau)$
to be regarded as a delta function,
we can obtain
\begin{align}
 \tilde\Psi(\Omega_a, \Omega_b) \propto \sqrt{\frac{T}{\tau}}
 \frac{1}{ \Delta}, \quad P_2 \propto \frac{T}{\tau} \frac{1}{\Delta^2},
 \label{Eq:two_gen}
\end{align}
where $\Delta \equiv (\Omega_a-\Omega_b)/2 - \bar\omega_-$
corresponds to the detuning from the intermediate state.
It should be noted that the two-photon absorption probability scales
as $P_2 \propto 1/\Delta^2$.
This result is consistent with the fact that
two-photon processes via a virtual state 
are a function of  $1/\Delta^2$,
which is derived for two cavity modes \cite{Loudon}.
Equation (\ref{Eq:two_gen}) also indicates that
the time correlation $\tau \ll T$,
which is a unique property of time-correlated photon pairs,
substantially enhances the two-photon absorption probability $P_2$.

\begin{figure}[]
\begin{center}
 \includegraphics[width=7cm]{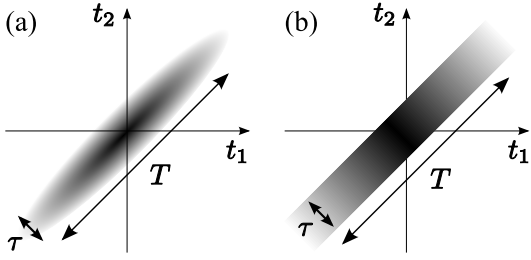}
 \caption{Two-photon wavefunction: (a) Gaussian 
and (b) rectangular.}
 \label{Fig:wf}
\end{center}
\end{figure}

Next, we compare the probability   of  two-photon excitation, $P_2$,
with that of one-photon excitation to the intermediate state $\ket{a}$,
$P_1$,
for the incidence of a photon pair.
For any two-photon wavepacket,
we can calculate $P_2$  
and $P_1$ using eq.~(\ref{Eq:P2}) 
and eq.~(\ref{Eq:P1}), respectively.
The probabilities $P_1$ and $P_2$ are significantly affected by
the shape of the two-photon wavefunction,
as described in Sec.~\ref{Sec:Theory}.
In the following subsections, we  investigate these probabilities
in two cases:
the case of a Gaussian wavefunction, as shown in 
Fig.~\ref{Fig:wf}(a),
and the case of a rectangular wavefunction, as shown in 
Fig.~\ref{Fig:wf}(b).
The Gaussian wavefunction can be obtained by restricting the spectrum
of photon pairs using a Gaussian filter.
The rectangular wavefunction can be generated by
a photon pairs via type-II spontaneous parametric-down conversion 
due to the difference in group velocities of two photons \cite{Shih:IEEE03,
PhysRevA.42.4102}.

\subsection{Gaussian wavefunction\label{type1}}

The Gaussian two-photon wavefunction is represented as
\begin{align}
 \psi(t_1, t_2) = a_1 & \ee^{- (t_1+t_2)^2 / 16 T^2} 
 \ee^{- (t_1-t_2)^2 / 4 \tau^2} \nonumber \\
 & \times \ee^{-\ii \bar{\omega}_1 t_1}
 \ee^{-\ii \bar{\omega}_2 t_2}, \label{Eq:psi_time1}
\end{align}
where $\bar{\omega}_1$ ($\bar{\omega}_2$) is the central frequency
of the signal (idler) photon and 
$a_1=1/\sqrt{2 \pi \tau T}$ is the normalization factor.
The Fourier transform of this wavefunction is easily derived as
\begin{align}
 \Psi(\omega_1, \omega_2) = A_1 \ee^{- (\omega_+
 -\bar{\omega}_+)^2  T^2} \,
 \ee^{- (\omega_- -\bar{\omega}_-)^2 \tau^2},
\label{Eq:typeI-1}
\end{align}
where $A_1=\sqrt{2 \tau T / \pi}$.
From its definition given in eq.~(\ref{Eq:Psi_t}), we obtain
\begin{align}
 \tilde{\Psi} (\omega_1, \omega_2) = \frac{A_1}{2} \ee^{- (\omega_+
 -\bar{\omega}_+)^2  T^2} \, 
 F\left( (\omega_- - \bar{\omega}_-) \tau \right), 
\label{Eq:typeI-2}
\end{align}
where $F(\xi)$ is called plasma dispersion function \cite{Fried} and
defined as
\begin{align}
 F(\xi) = \ee^{-\xi^2} \left(
 1 + \frac{2 \ii}{\sqrt{\pi}} \int^\xi_0 \ee^{y^2} \dd y
 \right).
\end{align}
It should be noted
 that the Gauss function in the last factor of eq.~(\ref{Eq:typeI-1})
is replaced with the plasma dispersion function
in  eq.~(\ref{Eq:typeI-2}).

By substituting eq.~(\ref{Eq:typeI-2}) in eq.~(\ref{Eq:P2})
and assuming that the sum of the two-photon frequency is tuned to
the two-photon transition, i.e., $\bar{\omega}_+
= \Omega_a+\Omega_b$,
we obtain the probability of the two-photon transition as
\begin{align}
 P_2 = \pi^2 r_1 r_2 A_1^2 \, |F(\Delta \cdot \tau)|^2.
\end{align}
From eqs.~(\ref{Eq:typeI-1}) and (\ref{Eq:P2}),
the probability of the one-photon transition
to the intermediate state $\ket{a}$ can be derived as
\begin{align}
 P_1 &= 2 \pi r_1 A_1^2
 \infint \dd \omega_2  \nonumber \\
 & \hspace{1cm} \times \ee^{-2 (\omega_2 - \Omega_b)^2 T^2} 
 \, \ee^{-(2 \Delta -\omega_2 + \Omega_b)^2 \tau^2 /2 }
 \nonumber \\
 &=\frac{\sqrt{2}\pi^{3/2} r_1 A_1^2}{T} \ee^{-2 (\Delta \cdot \tau)^2}
\end{align}
In the above derivation, we assume that $T \gg \tau$,
which allows us to consider $\ee^{-2 (\omega_2-\Omega_b)^2T^2}$
as the delta function $\sqrt{ \pi/2} \, \delta(\omega_2-\Omega_b) /T$.

\begin{figure}[]
 \begin{center}
 \psfrag{f1}[c][c]{$ |F(\Delta \cdot \tau)|^2, \ \ee^{-2(\Delta \cdot \tau)^2}$}

 \psfrag{f2}[c][c]{$ {\rm sinc}^2 (\Delta \cdot \tau/2), \ 
{\rm sinc}^2 (\Delta \cdot \tau)$}
 \psfrag{time}[c][c]{ correlation time $\Delta \cdot \tau$}
  \includegraphics[width=5cm]{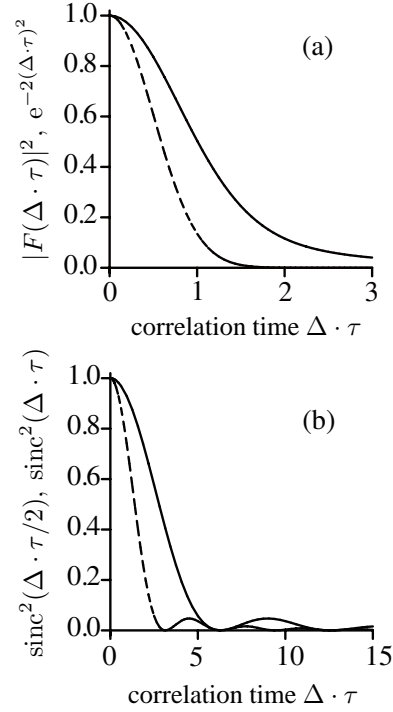}
   \caption{(a) Comparison of
  $|F(\Delta \cdot \tau)|^2$ (solid line) with
  $\ee^{-2 (\Delta \cdot \tau)^2}$ (dashed line).
(b) Comparison of ${\rm sinc}^2 (\Delta \cdot \tau/2)$ (solid line) with
${\rm sinc}^2 (\Delta \cdot \tau)$ (dashed line).}
\label{Fig:func}
 \end{center}
\end{figure}

Here, 
we introduce the ratio
\begin{align}
 R_G  \equiv \frac{P_2}{P_1} =
 \sqrt{\frac{\pi}{2}} r_2 T \,  \cdot \, \frac{|F(\Delta \cdot \tau)|^2}
 {\ee^{-2 (\Delta \cdot \tau)^2}}.\label{Eq:ratio1}
\end{align}
In the second factor, which is a function of the correlation time $\tau$,
both the Gauss function $\ee^{-2 (\Delta \cdot \tau)^2}$ in the numerator
and the square of the plasma dispersion function
$|F(\Delta \cdot \tau)|^2$ in the denominator
are monotonously decreasing functions, as shown in Fig.~\ref{Fig:func}(a).
However, there exists a critical difference
in the asymptotic behavior of $\Delta \cdot \tau > 1$:
$\ee^{-2 (\Delta \cdot \tau)^2}$ decreases rapidly,
while $|F(\Delta \cdot \tau)|^2$ decreases
 slowly as $1/(\Delta \tau)^2$.
Hence, we expect that the two-photon transition probability
$P_2$ exceeds the one-photon
transition probability $P_1$ depending on the value of $\Delta$ and
selected $\tau$,
even if the first factor in eq.~(\ref{Eq:ratio1}) is less than 1.

\subsection{Rectangular wavefunction\label{type2}}

By performing the type-II spontaneous parametric-down conversion
and using a birefringent crystal to achieve group velocity compensation,
we  prepare the two-photon wavefunction, which is expressed  as
\begin{align}
 \psi(t_1, t_2) = a_2 & \, \ee^{- (t_1+t_2)^2 / 16 T^2} \nonumber \\
 & \times \Pi_\tau(t_1-t_2) \ee^{-\ii \bar{\omega}_1 t_1} \,
 \ee^{-\ii \bar{\omega}_2 t_2}, \label{Eq:psi_time2}
\end{align}
where $\Pi_\tau(t)$ is a window function defined as
\begin{align}
 \Pi_\tau(t) = 
\begin{cases}
 0 & (|t|>\tau) ,\\
 1  & (|t|<\tau),
\end{cases}
\end{align}
and the normalization constant is $a_2=1/(8 \pi)^{1/4}/ \sqrt{\tau T}$
\cite{Shih:IEEE03,PhysRevA.50.5122}.

As shown in Sec.~\ref{type1},
we can deduce $\Psi(\omega_1, \omega_2)$ and $\tilde{\Psi}
(\omega_1, \omega_2)$ as follows:
\begin{align}
 \Psi(\omega_1, \omega_2) &= A_2 \,
 \ee^{-(\omega_+ -\bar\omega_+)T^2} \,
 {\rm sinc} \left( (\omega_- - \bar\omega_-)
 \tau \right) \label{Eq:psi-2} \\
 \tilde\Psi(\omega_1,  \omega_2) &= \frac{A_2}{2} \,
 \ee^{-(\omega_+ -\bar\omega_+)T^2} \, \nonumber \\
 & \hspace{-5mm}\times {\rm sinc} \left( \frac{\omega_- - \bar\omega_-}{2}
 \tau \right)
 \ee^{\ii(\omega_--\bar\omega_-)\tau/2}, \label{Eq:psi-t2}
\end{align} 
where ${\rm sinc}\,x \equiv \sin  x/x$ and $A_2=(2/\pi^3)^{1/4}\sqrt{\tau T}$.
On substituting eqs.~(\ref{Eq:psi-2}) and (\ref{Eq:psi-t2})
in eqs.~(\ref{Eq:P2}) and (\ref{Eq:P1}), respectively, we obtain
\begin{align}
 P_1 &= \frac{\sqrt{2}\pi^{3/2}r_1 A_2^2}{T} \, {\rm sinc^2}
 \left( \Delta \cdot \tau \right),\\
 P_2 &= \pi^2 r_1 r_2 A_2^2 \, {\rm sinc^2}
 \left( \frac{\Delta \cdot \tau}{2} \right).
\end{align}
Then, the ratio $R_r$ is given as
\begin{align}
  R_r \equiv \frac{P_2}{P_1} =
 \sqrt{\frac{\pi}{2}} r_2 T \,  \cdot \, \frac{{\rm sinc^2}\left(
 \frac{\Delta \cdot \tau}{2} \right)} 
 {{\rm sinc^2} (\Delta \cdot \tau)}.\label{Eq:ratio2}
\end{align}
The first factor in this equation 
is identical to that in eq.~(\ref{Eq:ratio1})
because the two wavefunctions expressed in eqs.~(\ref{Eq:psi_time1})
and (\ref{Eq:psi_time2}) exhibit the same function with respect to $t_1+t_2$.
In the second factor, the period of the function of the numerator is
twice as long as that of the denominator, as shown in Fig.~\ref{Fig:func}(b).
For $\tau=2 \pi n / \Delta$, where $n$ is an integer,
the two-photon absorption is no longer observed.
Fei {\it et al.}  predicted this phenomenon 
and called it
 {\it entanglement-induced two-photon transparency} \cite{fei:1679}.
From this study, we have found that
for $\tau=\pi (2 n +1)/\Delta$,
 only the two-photon absorption is induced,
while the one-photon absorption is completely suppressed.

\section{Conclusion}

We derived the probabilities for one-photon  
and two-photon transitions 
($P_1$ and $P_2$, respectively)
for  two-photon
states in general.
We showed that the two-photon absorption can be dramatically enhanced 
because of the time correlation $\tau \ll T$ of general two-photon 
wavefunctions of a photon pair.
Then, we dealt with two typical examples of a Gaussian wavefunction and
a rectangular wavefunction and calculated $P_1$ and $P_2$
in both cases.
The probabilities $P_1$ and $P_2$ were found to behave differently
with respect to $\Delta \cdot \tau$.
On the basis of this difference in behavior,
we can enhance the two-photon absorption while suppressing
the undesired one-photon absorption, by adjusting the detuning $\Delta$
and the correlation time $\tau$.
In particular,
the photon pair having a rectangular wavefuction 
can induce the two-photon transition without any one-photon loss
under certain special conditions.

\begin{acknowledgements}
This study is supported by the Global COE Program 
``Photonics and Electronics Science and Engineering'' at 
Kyoto University.
\end{acknowledgements}

\end{document}